\documentclass[a4paper,fleqn,usenatbib]{mnras}
\usepackage{newtxtext,newtxmath}
\usepackage[T1]{fontenc}
\usepackage{ae,aecompl}

\usepackage{float}
\usepackage{array}
\usepackage{gensymb}
\usepackage{pdflscape}	
\usepackage{graphicx}	
\usepackage{amsmath}	
\graphicspath{{./figures/}}

\title[T. G. Nguyen et al. (2020)]{Modelling the atmosphere of lava planet K2-141b: implications for low and high resolution spectroscopy}

\author[T. G. Nguyen et al.]{
T. Giang Nguyen,$^{1}$\thanks{E-mail: giang@yorku.ca}
Nicolas B. Cowan,$^{2}$
Agnibha Banerjee$^{3}$
and John E. Moores$^{1}$
\\
$^{1}$Centre for Research in Earth and Space Sciences, York University, 4700 Keele st., Toronto, Ontario, M3J 1P3, Canada\\
$^{2}$Department of Earth and Planetary Sciences, and Department of Physics, McGill University, 3550 Rue University, Montr\'eal, Qu\'ebec, H3A 2A7, Canada\\
$^{3}$Department of Physical Sciences, Indian Institute of Science Education and Research, Kolkata, Mohanpur, 741246, India}

\date{Accepted XXX. Received YYY; in original form ZZZ}

\pubyear{2020}

\begin{document}
\label{firstpage}
\pagerange{\pageref{firstpage}--\pageref{lastpage}}
\maketitle

\begin{abstract}
Transit searches have uncovered Earth-size planets orbiting so close to their host star that their surface should be molten, so-called lava planets. We present idealized simulations of the atmosphere of lava planet K2-141b and calculate the return flow of material via circulation in the magma ocean. We then compare how pure Na, SiO, or SiO$_2$ atmospheres would impact future observations. The more volatile Na atmosphere is thickest followed by SiO and SiO$_2$, as expected. Despite its low vapour pressure, we find that a SiO$_2$ atmosphere is easier to observe via transit spectroscopy due to its greater scale height near the day-night terminator and the planetary radial velocity and acceleration are very high, facilitating high dispersion spectroscopy. The special geometry that arises from very small orbits allows for a wide range of limb observations for K2-141b. After determining the magma ocean depth, we infer that the ocean circulation required for SiO steady-state flow is only $10^{-4}$ m/s while the equivalent return flow for Na is several orders of magnitude greater. This suggests that a steady-state Na atmosphere cannot be sustained and that the surface will evolve over time.
\end{abstract}




\section{Introduction}

Lava planets are a class of rocky exoplanets that orbit so close to their star that parts of their surface are molten (for a review of ultra-short-period planets, see \citealp{winn2018kepler}). Indeed, dayside temperatures can be hot enough to maintain a rock vapour atmosphere detectable through transit and eclipse spectroscopy, as well as phase curves \citep{schaefer2009chemistry}. Theoretical studies of lava planets and their relatively volatile sodium atmospheres have been published for CoRot-7b \citep{leger2011extreme}, Kepler-10b, 55 Cnc-e \citep{castan2011atmospheres}, KIC 12556548, and HD 219134b \citep{kite2016atmosphere}. These studies utilized a 1D model that solves for the flow of mass, momentum, and energy from the hot dayside to the cold nightside.

We study K2-141b, the highest signal-to-noise lava planet discovered to date, as its phase variations have been measured by K2 \citep{malavolta2018ultra,barragan2018k2}, Spitzer \citep{kreidberg2019taking}, and possibly in the future with the James Webb Space Telescope (JWST) (e.g \citealp{stevenson2016transiting}). We use essentially the same model as the above studies, but we employ more accurate surface temperature calculations. We consider three compositional end-members for the atmosphere: 1) pure sodium (Na), due to its high volatility, 2) silicon monoxide (SiO), or 3) pure silica (SiO$_2$) due to their abundance in the crust of rocky planets. These three cases bracket the range of plausible atmospheric compositions \citep{schaefer2009chemistry}. We focus on the mass exchange between surface and atmosphere to provide insights on the magma ocean return flow. We also use the results to anticipate future observations with space telescopes as well as ground-based high-dispersion spectroscopy.

\section{Theory}

\subsection{The surface and below}

\begin{figure}
	\centering	
	\includegraphics[width=\columnwidth,height=\textheight,keepaspectratio]{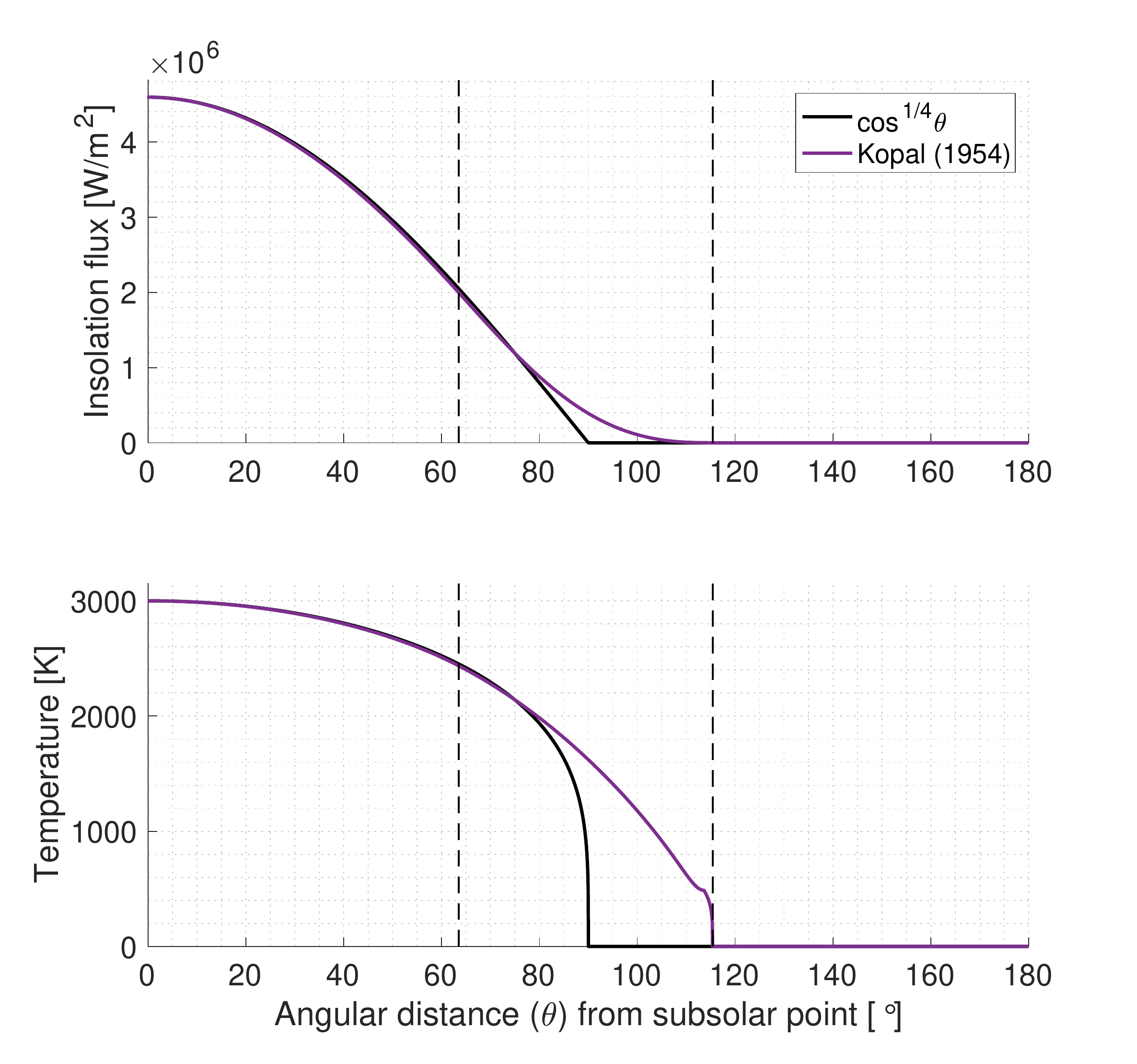}
	\caption{\textit{Top: the illumination received by K2-141b as a function of angular distance from the sub-stellar point. Bottom: equilibrium temperature. In both panels, the black line shows the cos\textsuperscript{1/4}$\theta$ flux dependence adopted by previous researchers while the purple line shows more accurate formulation that accounts for the large angular size of the star as seen from the planet.}}
	\label{fig_surftem}	
\end{figure}

We adopt planetary parameters from \cite{malavolta2018ultra} and \cite{barragan2018k2}. The subsolar equilibrium temperature (the planet's irradiation temperature) is 3038 $\pm$ 64 K; we adopt the round number of 3000 K for the remainder of this study. Previous estimates of the surface temperature ($T_s$) by \cite{castan2011atmospheres} and \cite{kite2016atmosphere} accounted for the zero-albedo local radiative equilibrium temperature and geothermal heating:

\begin{equation}
T_s = 
\begin{cases}
(T_{ss} - T_{as})\cos^{1/4}(\theta) + T_{as} & \text{for $\theta \leq 90\degree$,} \\
T_{as} & \text{for $\theta > 90\degree$,}
\end{cases}
\end{equation}

\noindent where $T_{ss}$ and $T_{as}$ are the temperature at the subsolar and antisolar point, respectively, and $\theta$ is the angular distance from the subsolar point. The $\cos^{1/4}\theta$ expression, however, is only valid in the limit of a distant star, $R_*/a << 1$. K2-141b has such a tight orbit that the star has an angular diameter of 50 degrees as seen from the planet and the surface can still be illuminated well past $\theta = 90\degree$. Here, we employ a more accurate analytic expression for the surface temperature originally developed for binary star systems \citep{kopal1954photometric}. The surface temperature of K2-141b calculated using these two schemes is shown in Fig. \ref{fig_surftem}. Beyond $\theta \simeq 115\degree$, the illumination flux is so low that the temperature is mainly dependent on a geothermal flux, which we neglect. Our model will show that the atmosphere fully condenses out well before this point.

With the surface temperature now calculated, we can infer physical properties of the magma ocean by making a few assumptions and simplifications. As silicate materials dominate the composition of rocky planets \citep{schaefer2009chemistry} and K2-141b's bulk density is comparable to that of the Earth \citep{malavolta2018ultra}, we assume that K2-141b's crust is entirely silica (SiO$_2$). We further assume that the temperature is vertically constant below the surface for the top 150 km. We neglect geothermal heating, which is reasonable if the ocean is relatively shallow and vertically well mixed.

To determine the magma ocean depth, we combine the planet's known surface gravity (21.8 m/s$^2$) with the density of SiO$_2$ (2650 kg/m$^3$) to obtain the pressure as a function of depth. These calculated pressures and temperatures then allow us to predict the phase (liquid vs solid) of the SiO$_2$ as a function of depth using the phase diagram shown in Fig. \ref{fig_phasediagram} \citep{swamy1994thermodynamic}. Using these data, we calculate the depth of magma ocean as a function of the surface temperature at the top of the column and hence as a function of $\theta$,  also shown in Fig. \ref{fig_phasediagram}. In Section \ref{sec_oce} we will close the loop between the atmosphere and the surface: mass must be redistributed via ocean currents to account for the constant evaporation and condensation.

\subsection{The atmosphere}

As a short-period planet on a circular orbit, K2-141b is expected to be tidally locked into synchronous rotation, leading to a permanent dayside and nightside. Not only does this introduce symmetry in how the planet is illuminated, it also leads to equilibrium dynamics \citep{schaefer2009chemistry}. The atmosphere of a lava planet originates from the sublimation/evaporation of the surface and the flow is maintained by the pressure gradient between the nightside and dayside. The solutions are therefore dependent on the surface temperature and solely a function of the angular distance from the sub-solar point ($\theta_{ss}$).

We assume a hydrostatically bound ``thin" atmosphere with negligible absorption of radiation. Although radiative transfer simulations of mineral-vapour atmospheres suggest that this is not the case \citep{ito2015theoretical}, it is a useful first approximation and makes the problem tractable. We also neglect Coriolis forces to keep the problem one-dimensional although our estimated Rossby number of $10^{-1}$ suggests that Coriolis forces may be stronger than on other lava planets \citep{castan2011atmospheres}. Lastly, we assume that the atmosphere is turbulent and well-mixed, leading the vertically-uniform wind velocity, $V$. Other state variables, pressure $P$ and temperature $T$, are values calculated at the top of the boundary layer \citep{ingersoll1985supersonic}. We therefore use the shallow-water equations, commonly used for hydrodynamical calculations of non-linear evaporation/sublimation processes (e.g \citealp{ingersoll1985supersonic, ding2018global}).

We compute the surface temperature ahead of time because, as stated above, we assume it is unaffected by the atmosphere. The atmosphere is idealized to be pure and condensible because of the model's limits. We first consider a pure Na atmosphere, neglecting dissociation processes that unbind Na from other molecules as is done by previous studies of lava planets (e.g., \citealp{castan2011atmospheres,kite2016atmosphere}). Na is relatively scarce in a rocky planet's crust \citep{schaefer2009chemistry} whereas SiO$_2$ is the dominant constituent. Therefore, we also run the model using SiO in the same spirit as Na. Since gas-phase SiO will likely condense as SiO$_2$ and may recombine with oxygen even before condensing, we finally use a pure SiO$_2$ atmosphere.

The model was derived from the shallow-water equations by \cite{ingersoll1985supersonic} and adapted by \cite{castan2011atmospheres} for exoplanets. It describes the conservation of mass (Eq. \ref{eq_mass}), momentum (Eq. \ref{eq_mom}) and energy (Eq. \ref{eq_en}). Mass conservation is as follows: 

\begin{equation}
\partial_t h + \nabla\cdot(\rho hv) = 0,
\end{equation}

\noindent where $\rho$ is the fluid density, $h$ is the fluid's column thickness, and $v$ is the horizontal flow. Applying this equation to our case requires time invariance, $\partial_t h$ = 0, hydrostatic equilibrium, $h = P/(\rho g)$, and a source/sink to account for evaporation/condensation such that the right hand side of Equation (2) is no longer 0 but some mass flux rate,  $E$ [molecules s$^{-1}$ m$^{-2}$]. The final step is to change to spherical coordinates which gives our mass conservation equation:

\begin{equation}
\frac{1}{r \ \sin(\theta)}\frac{d}{d\theta}\bigg(\frac{V \ P \ \sin(\theta)}{g}\bigg) = m \ E,
\label{eq_mass}
\end{equation}

\noindent where $r$ is the planet's radius ($9.62\times10^6$ m), $m$ is the mass per molecule, and $g$ is the surface gravity ($21.8$ m/s$^2$). We can do the same for momentum, which is affected by pressure gradient forces and surface drag $\tau$ [ kg m$^{-1}$ s$^{-2}$]. This relation has the general form:

\begin{equation}
\partial_t (\rho v) + \nabla \cdot (\rho v^2) = -\nabla P + \tau.
\end{equation}

A steady state flow means that $\partial_t (\rho v)$ is 0. Substituting the mass for a product of $V$, $P$ and $g$, integrating $\nabla P$ vertically, collecting like terms, and changing to spherical coordinates yields the momentum conservation equation:

\begin{equation}
\frac{1}{r \ \sin(\theta)}\frac{d}{d\theta}\bigg(\frac{(V^2 + \beta \ C_p \ T)P \ \sin(\theta)}{g}\bigg) = \frac{\beta \ C_p \ T \ P}{g \ r \ \tan(\theta)} + \tau,
\label{eq_mom}
\end{equation}

\noindent where $C_P$ is the heat capacity, $\beta$ is the ratio $R/(R+C_p)$, $R$ is the gas constant. Parameters specific to each atmospheric constituent are included in Appendix \ref{App_Gas}. Energy is conserved if the vertically integrated kinetic energy flux, internal heat energy, gravitational energy, and work done via adiabatic expansion is balanced out by energy exchanged with the surface, $Q$ [W m$^{-2}$]. The sum of the internal heat, gravity, and work terms becomes the enthalpy per unit mass at the top of the boundary layer: $C_p T$. Considering 1D steady state flow in spherical coordinates yields the equation for energy conservation:

\begin{equation}
\frac{1}{r \ \sin(\theta)}\frac{d}{d\theta}\bigg(\frac{(V^2/2 + \beta \ C_p \ T)V \ P \ \sin(\theta)}{g}\bigg) = Q.
\label{eq_en}
\end{equation}

Calculations of fluxes $E$, $\tau$, and $Q$ are included in Appendix \ref{App_flux}. Detailed derivation and steps shown between the generalized equations and our conservation equations can be found in \cite{ingersoll1985supersonic}. As the system only depends on $\theta$ through the state variables $P$ [Pa], $V$ [m/s], and $T$ [K], the equilibrium solution is obtained by iteratively integrating the fluxes. This computationally solves the ordinary differential equations as an initial boundary condition problem (details in Appendix \ref{App_solve}).

\section{Results}

\subsection{The atmosphere}

\begin{figure}
	\centering	
	\includegraphics[width=\columnwidth,height=\textheight,keepaspectratio]{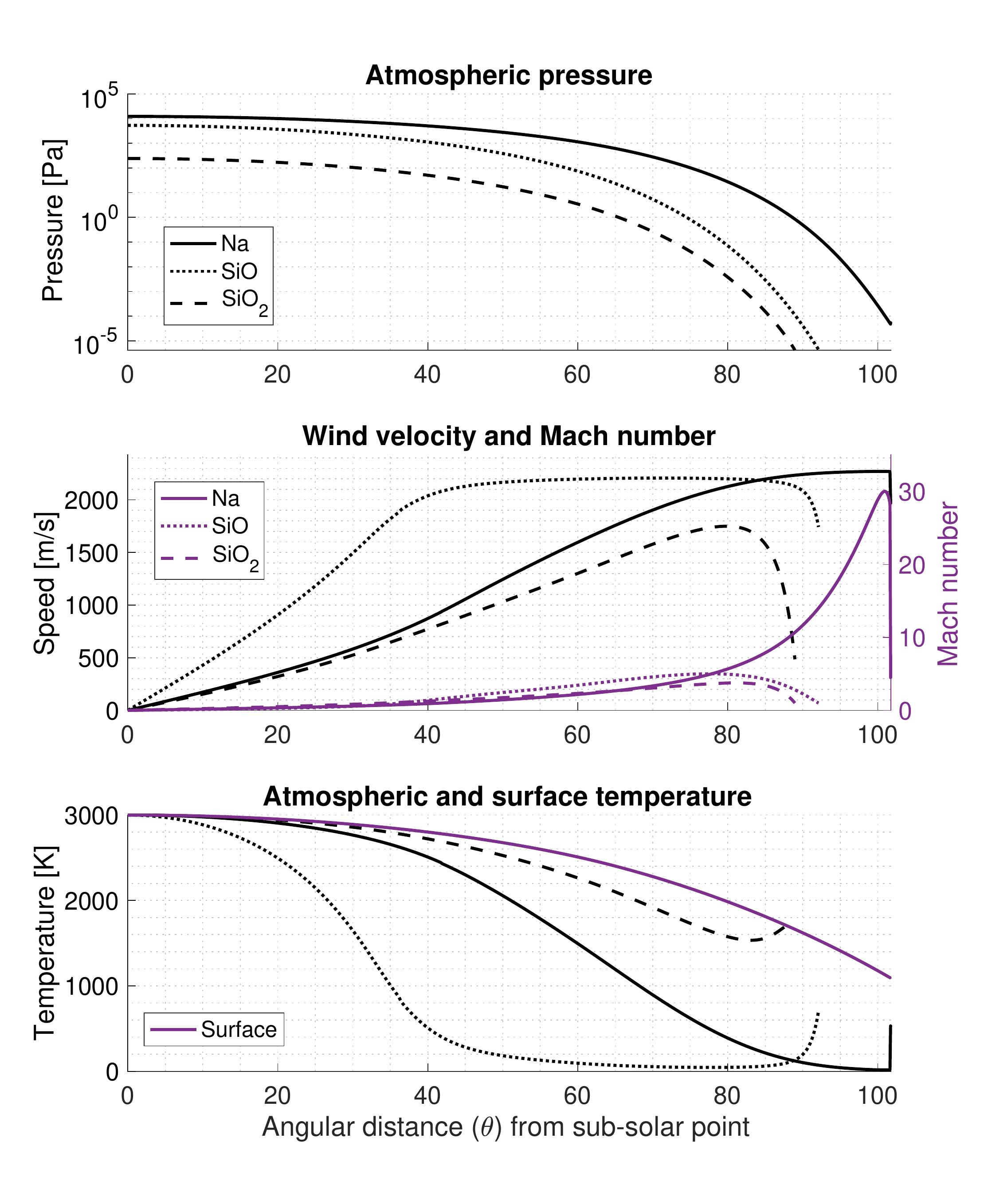}
	\caption{Top panel: atmospheric boundary-layer pressure for a pure sodium (Na), silicon monoxide (SiO), and silica (SiO$_2$) atmosphere. Middle panel: wind velocity (left/black) and mach number (right/purple). Bottom panel: atmospheric and surface temperatures (solid purple line).}
	\label{fig_PVT}
\end{figure}

We used the model described above to determine the steady-state flow of a pure Na, SiO, and SiO$_2$ atmosphere for lava planet K2-141b. Atmospheric pressure, wind velocity, and temperature for the three cases are shown in Fig. \ref{fig_PVT}. Na is the most volatile molecule, making the Na atmosphere the thickest of the three with the maximum pressure at the subsolar point of 13.8 kPa. The Na atmosphere fully condenses out at $\theta = 102\degree$ where wind speed can reach up to 2.3 km/s (Mach 30). The atmospheric temperature of Na can drop to near zero before fully collapsing.

SiO is not as volatile as Na making the SiO atmosphere thinner with a maximum pressure of 5.25 kPa. The wind, however, accelerates much faster than the Na case with velocities reaching up to 2.2 km/s and the atmosphere condenses out at $\theta = 92\degree$. The temperature also drops much faster. Running the model for SiO$_2$ produces the thinnest atmosphere with a maximum pressure of 240 Pa and collapses at $\theta = 89\degree$. The wind is slower with a maximum speed of 1.75 km/s but the SiO$_2$ atmosphere is much warmer as the temperature remains above 1500 K.

Evaporation and condensation rates vary significantly between the Na, SiO, and SiO$_2$ cases. Although Na and SiO have similar evaporation rates near the subsolar point, the SiO atmosphere starts to condense sooner than its Na counterpart. The evaporation rates for the three scenarios are shown in Fig. \ref{fig_OceVel}. 

\subsection{The magma ocean}
\label{sec_oce}

To maintain steady-state atmospheric flow, material must be transported back to the sub-stellar region along the surface, most likely through currents in the magma ocean. Horizontal mass transport is determined by calculating the cumulative mass that has evaporated and subsequently condensed. Therefore, the atmospheric mass transport ($M_T$) at some $\theta$ is given by: 

\begin{equation}
\label{eq_MT}
M_T(\theta) = \int_{0}^{\theta} m E(\theta) d\theta.
\end{equation}

The ocean return flow, $v_{o}$, that balances the atmospheric mass transport is therefore calculated via:

\begin{equation}
\label{eq_VO}
v_{o}(\theta) = M_T(\theta)/(\rho \ A(\theta)),
\end{equation}

\noindent where $\rho$ is the density of the magma ocean: 2650 kg/m$^3$ for SiO$_2$ or 2270 kg/m$^3$ for Na$_2$O (Na likely exists as Na$_2$O; \citealp{miguel2011compositions}). $A(\theta)$ is the cross-sectional area of the magma ocean given the magma ocean depth (estimated in section 2.1 and shown in Fig. \ref{fig_phasediagram}). Eqs. \ref{eq_MT} and \ref{eq_VO} predict the ocean velocity without accounting for the dissociation of oxygen. This approximation may be fine for the SiO$_2$ atmosphere, but not for the SiO and Na scenarios. To account for the oxygen dissociated from SiO$_2$ or Na$_2$O, we simply scale the $m$ value in Eq. \ref{eq_MT} to conserve the mass of oxygen during the evaporation and condensation processes: scaling factor = $(1+m_O/m_{SiO})$ for SiO and $(1+0.5 \times m_O/m_{Na})$ for Na.

\begin{figure}
	\centering	
	\includegraphics[width=\columnwidth,height=\textheight,keepaspectratio]{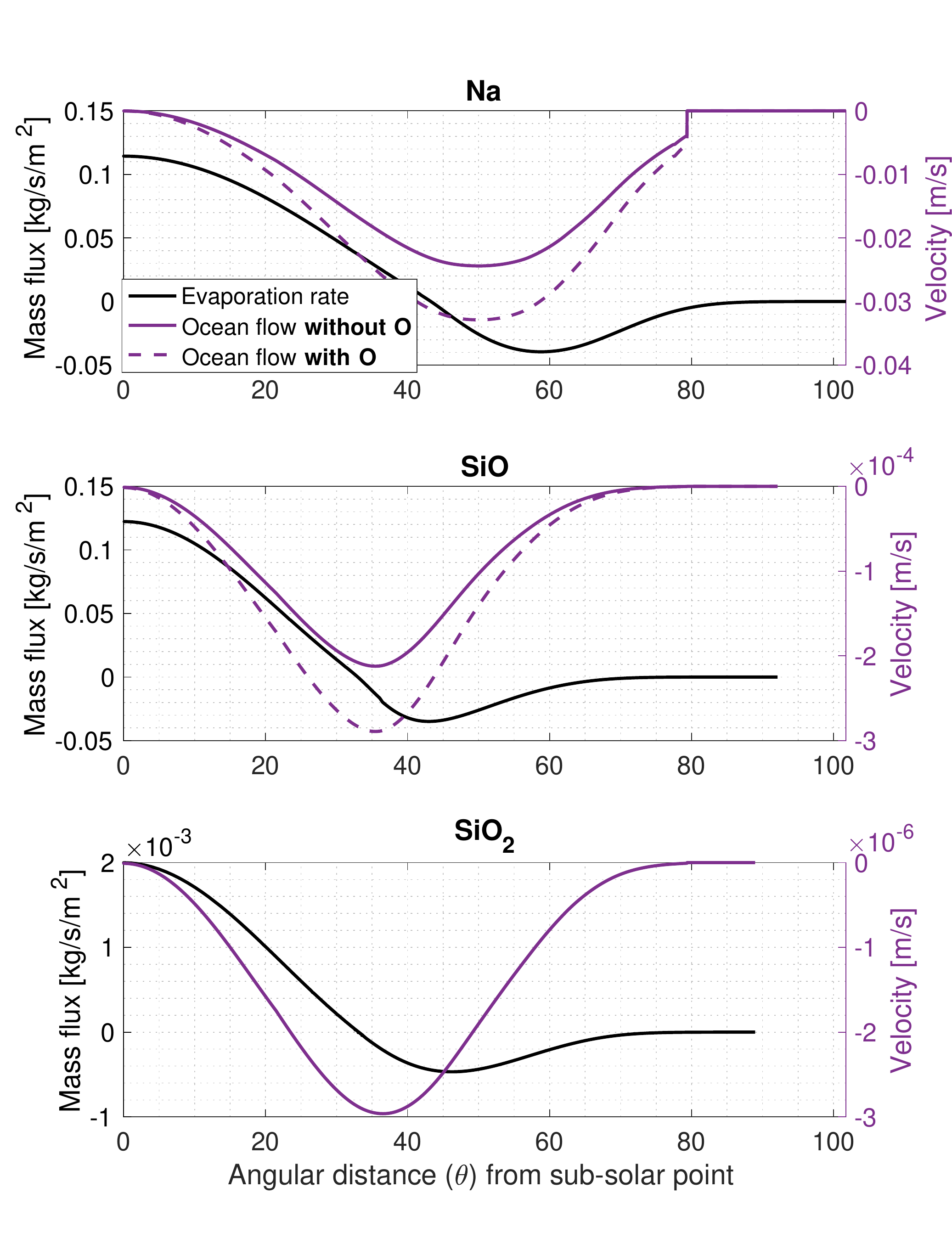}
	\caption{Left/black: evaporation rate per square metre (negative numbers denote condensation). Right/purple: velocity of magma ocean return flow. Negative values denote mass flow towards the subsolar point (in the negative $\theta$ direction). Solid purple lines denote calculations neglecting the transport of oxygen while dashed lines includes oxygen to maintain mass balance.}
	\label{fig_OceVel}	
\end{figure}

As the magma is predominantly SiO$_2$, the ocean circulation is dictated by our SiO atmosphere (recall that SiO has a much greater vapour pressure than SiO$_2$). These calculations yield a maximum magma ocean return flow of 2-3$\times10^{-4}$ m/s, very slow compared to ocean circulation on Earth. The evaporation rate and inferred ocean velocities are shown as Fig. \ref{fig_OceVel}.

If we assume a SiO$_2$ to Na ratio similar to that of a Bulk Silicate Earth (where Na$_2$O accounts for 0.7\% of the BSE), then Na accounts for 1.52\% by mass of our idealized crust \citep{miguel2011compositions}. Applying the same ratio to the magma ocean, we calculate the ocean circulation required to maintain mass balance for Na which resulted in a maximum velocity of  2-3 cm/s, two orders of magnitude faster than the expected circulation in the magma ocean. While not much mass condenses beyond the magma ocean for our SiO and SiO$_2$ cases, a significant amount of Na is deposited onto solid ground. Here, material can no longer be transported by ocean circulation, but rather by solid flow akin to glaciers flowing into the ocean on Earth.

\section{Discussion}
\subsection{Implications for observation}

We presented three scenarios where the atmosphere of K2-141b is either purely Na, SiO, or SiO$_2$ and modelled the pressure, flow, and temperature for each. While qualitatively similar, the results differ significantly depending on the atmospheric constituent, suggesting that observations could distinguish between the three scenarios. For example, the James Webb Space Telescope \citep{gardner2006james} could observe the near infrared (NIR) phase curve of K2-141b while the Hubble Space Telescope could observe it in the visible spectrum. Spectroscopy spanning the NIR SiO$_2$, IR SiO , or visible Na features could simultaneously probe the planetary surface and atmospheric temperature.

\begin{figure}
	\centering	
	\includegraphics[width=\columnwidth,height=\textheight,keepaspectratio]{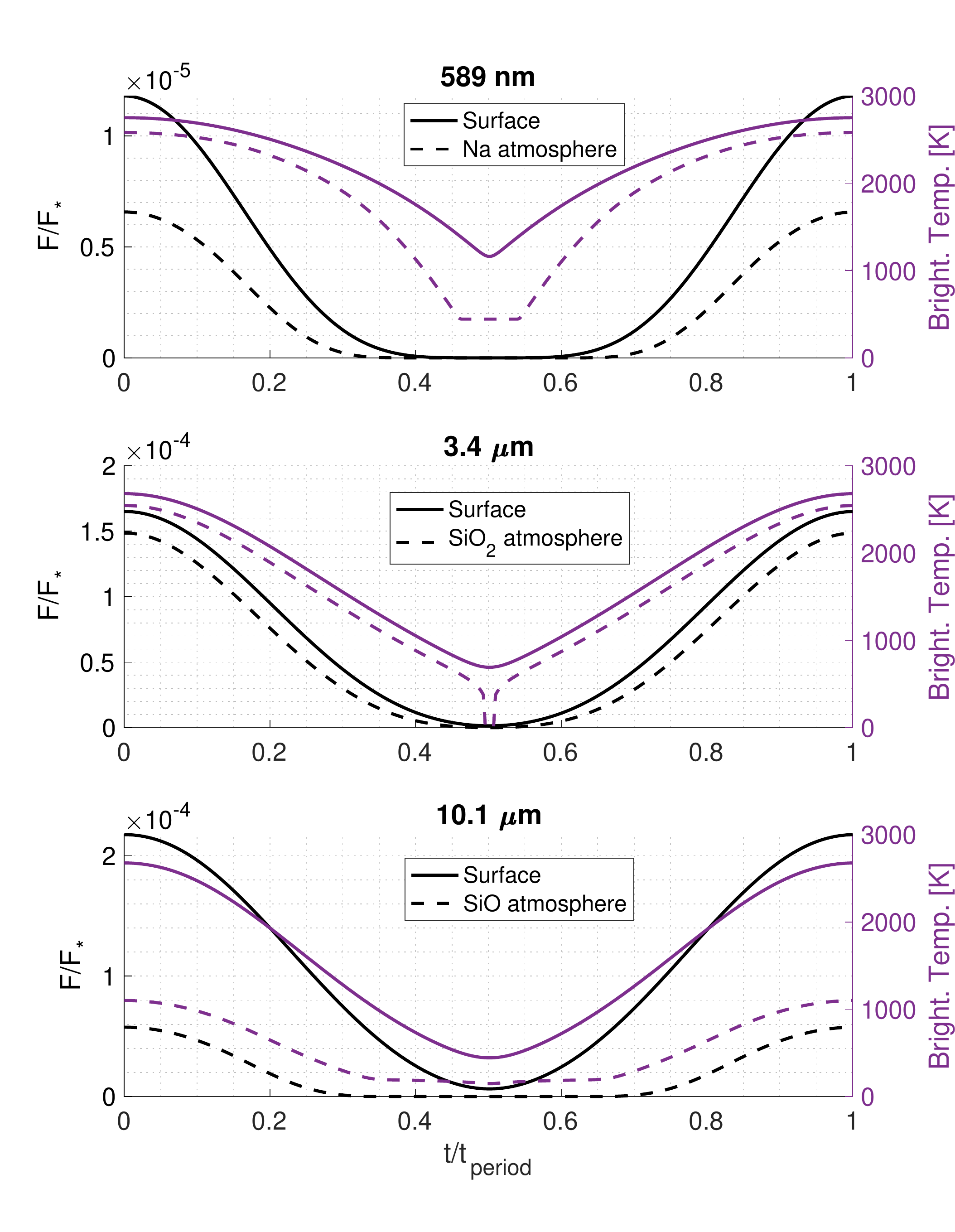}
	\caption[]{Top: predicted blackbody radiation of K2-141b at 589 Na spectral feature (left/black) and phase-dependent disk-integrated planetary brightness temperature (right/purple). Middle: the same for the 3.4 micron SiO$_2$ spectral feature. Bottom: the same for the 10.1 micron SiO spectral feature. Note that eclipses and transit have been omitted but would occur at 0 and 0.5 respectively. The phase curve is shown as the flux ratio between the planet and star. Solid lines refer to the surface (a wavelength just outside the spectral feature) and dashed lines refer to the atmosphere (a wavelength in the spectral feature).}
	\label{fig_emiss}
\end{figure}

In Fig. \ref{fig_emiss}, we show phase variations at wavelengths in and out of the 589 nm Na feature; we likewise show phase variations at wavelengths in and out of the 3.4 $\mu$m SiO$_2$ feature \citep{coblentz1977coblentz} and 10.1 $\mu$m SiO feature \citep{rawlings1968optical}. Phase variations are determined by calculating the expected wavelength-dependent blackbody radiative flux of the planet throughout an orbit. This process, detailed in Appendix \ref{App_obs}, involves transforming our 1D results to a 2D map, which we then integrate over the planet's visible hemisphere to get the expected flux measurements. 

As seen in Fig. \ref{fig_emiss}, the nightside flux and brightness temperature of both the surface and the Na and SiO atmospheres do not go to 0 at inferior conjunction (orbital phase of 0.5). This is because the visible hemisphere is still somewhat illuminated, and the Na and SiO atmospheres have not fully condensed. Despite being the thinnest, the SiO$_2$ atmosphere produces the highest radiative flux as it is significantly warmer than the other two. However, the SiO$_2$ atmosphere will have fully condensed out before  $\theta=90\degree$. This suggests that transit spectroscopy is impossible for a SiO$_2$ atmosphere, unlike Na and SiO, as the atmosphere is entirely hidden from view on the planet's dayside. We show below that this is not necessarily the case.

\begin{figure}
	\centering	
	\includegraphics[width=\columnwidth,height=\textheight,keepaspectratio]{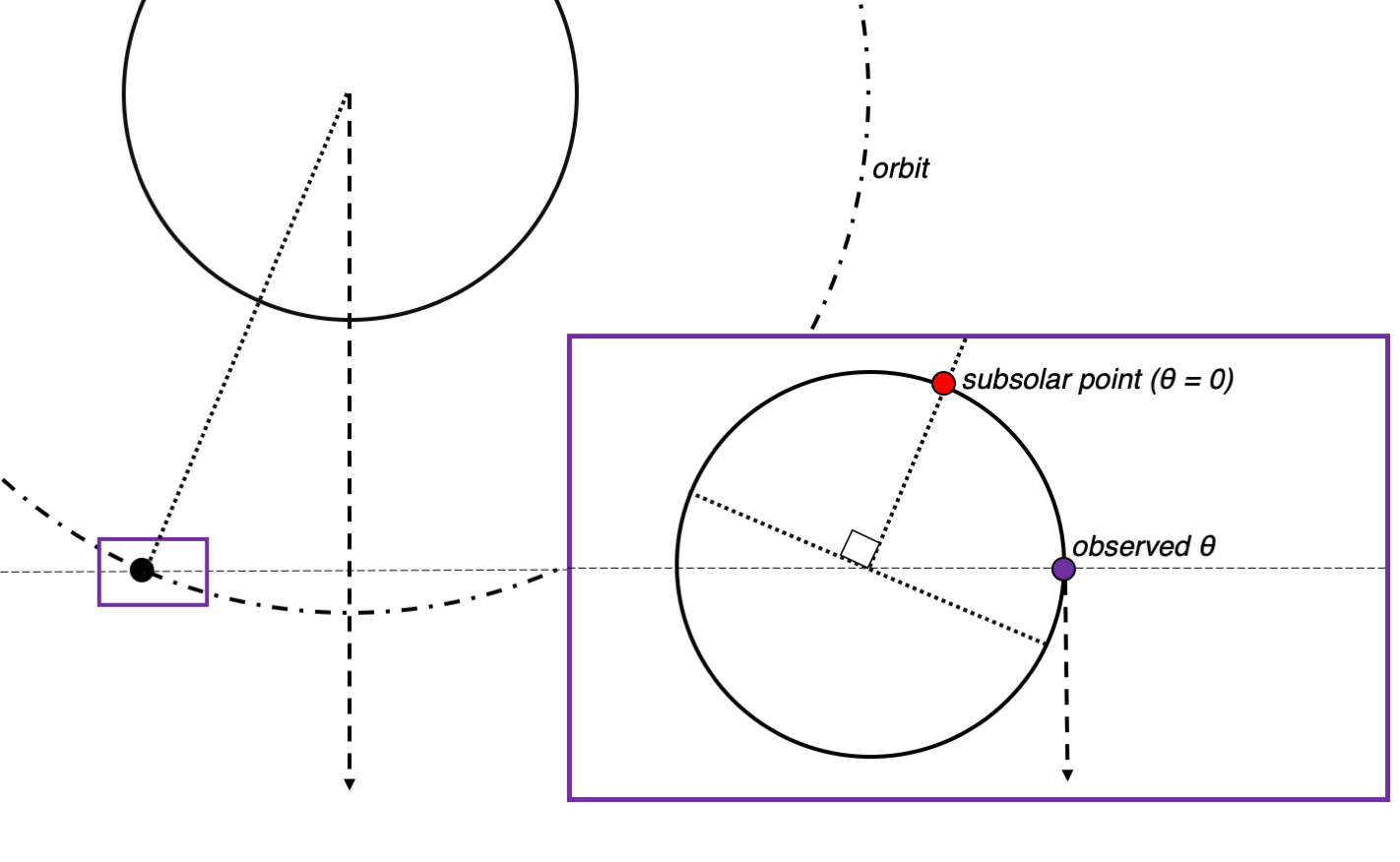}
	\caption{Illustration of the observed longitude during a transit (the observer is at the bottom of the page). The star and planetary orbit are drawn to scale.}
	\label{fig_transillus}
\end{figure}

In addition to its effect on illumination patterns and surface temperature, the especially close orbit of K2-141b has implications for transit geometry. It is generally assumed that transit spectroscopy only probes the atmosphere $90\degree$ from the subsolar point. As illustrated in Fig. \ref{fig_transillus}, regions probed during transit can be quite far from $\theta = 90\degree$ for ultra-short period planets. The start and end of transit therefore probe regions with a significant atmosphere even in the SiO$_2$ case. For the specific case of K2-141b and assuming an impact parameter of zero (consistent with observational constraints), ingress and egress probe equatorial regions at an angle $\tan^{-1}(R_*/a) \simeq 26\degree$ away from the usually assumed $\theta = 90\degree$ (for sight lines offset from the planetary equator, the observed $\theta$ is closer to $90\degree$).

\begin{figure}
	\centering	
	\includegraphics[width=\columnwidth,height=\textheight,keepaspectratio]{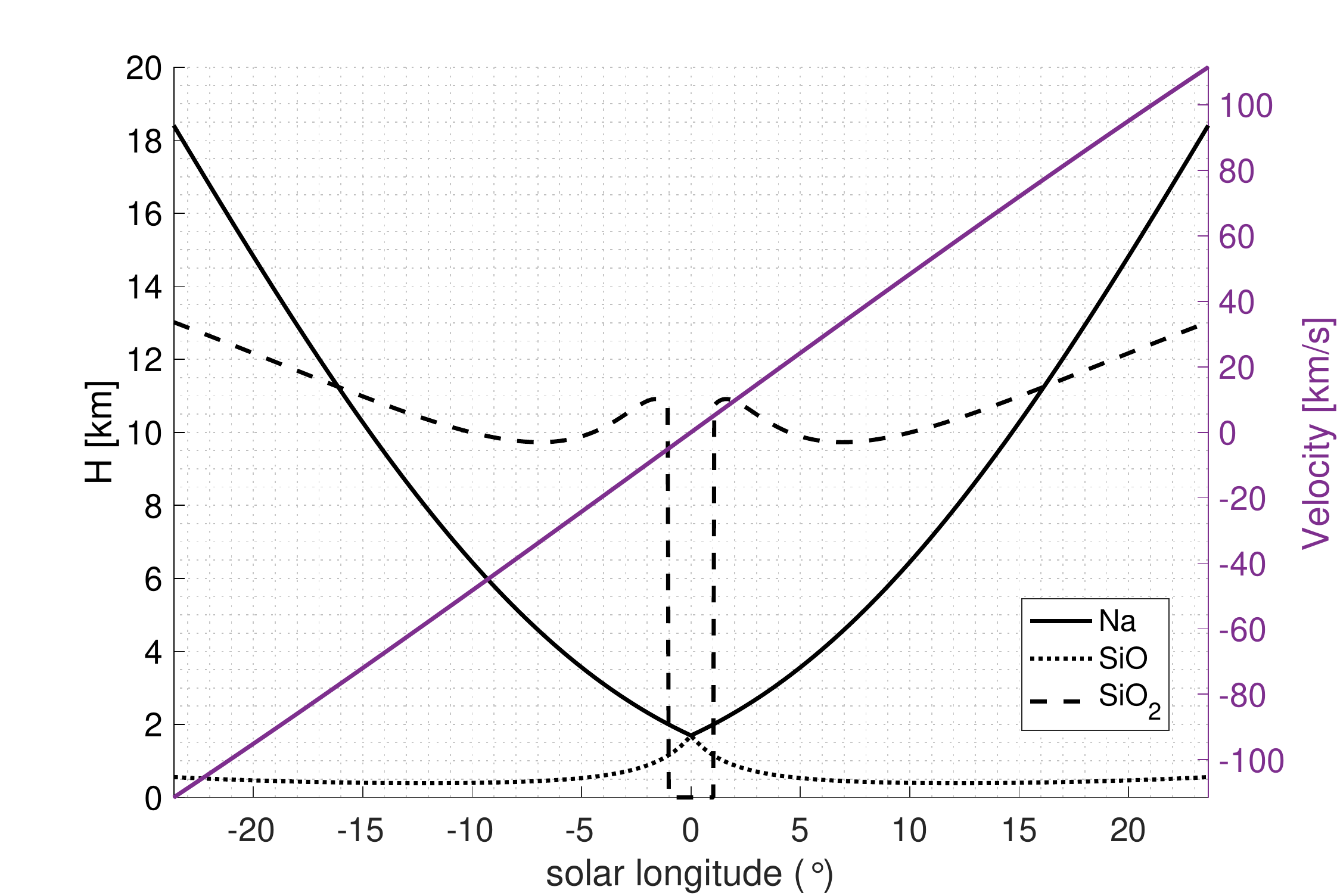}
	\caption{Anatomy of a transit of K2-141b, in degrees from transit. Left/black: observed scale height for the Na, SiO, and SiO$_2$ cases for regions along the equator. Even though the SiO$_2$ atmosphere collapses before $\theta=90\degree$ and hence is invisible in the centre of eclipse, the scale height of the SiO$_2$ atmosphere is bigger and observable for most of the planet's transit. Right/purple: the planet's velocity along the observer's line of sight (negative velocities denote motion towards the observer).  }
	\label{fig_transit}
\end{figure}

This opens up an avenue for high-dispersion spectroscopy where Doppler shifts from the changing radial velocity of the planet can help disentangle planetary and stellar absorption features and potentially detect atmospheric winds \citep{snellen2014high}. Note that the extreme orbit of K2-141b leads to significant changes in radial velocity over the course of the transit. The radial velocity and the observed scale height along the planetary equator are shown in Fig. \ref{fig_transit}.

If we assume the composition of K2-141b to be similar to the Bulk Silicate Earth, then there is 100x more silicate material than Na, by mass \citep{miguel2011compositions}. However, our results show that the mass transport required for a steady-state flow of Na is much greater than the $\simeq10^{-4}$ m/s provided by the SiO$_2$ magma ocean. This implies that a more realistic Na atmosphere is much thinner than what we predicted above because the steady-state flow cannot be sustained, leading to the ``evolved" state anticipated by \cite{kite2016atmosphere}. Also, any precipitation that falls beyond the magma ocean shore ($\theta>79\degree$) must be brought back to maintain mass conservation. This may occur via solid state flow analogous to glaciers on Earth, or via isostatic adjustment. If mass is transported back too slowly, the resulting mass imbalance could lead to reorientation of the planetary spin \citep{leconte2018continuous}.

\subsection{Model Limitations}

Our model makes promising predictions for observation of the lava planet K2-141b. However, our idealized approach to characterize K2-141b's atmosphere and magma ocean has limits which must be addressed. This subsection will discuss these limitations and list the caveats that come with our expected observation measurements as seen in Figures \ref{fig_emiss} and \ref{fig_transit}.

Our proposed pure or nearly pure SiO$_2$ magma ocean is likely affected by interior dynamics which we neglect. The calculated ocean flow only accounts for the bulk mass transport along $\theta$ to balance the evaporation and condensation rates. We also neglect Coriolis forces despite the fast rotation of K2-141b. Qualitatively, Coriolis force would deflect the wind en route to the cold nightside and the mass transport of both the atmosphere and the ocean would be constrained to a smaller area, presumably near the equator. Despite this, there should still be net mass transport in the atmosphere as the evaporated volatiles would still move from an area of high vapour pressure to areas of low vapour pressure. This also applies to the magma ocean as mass would need to recirculate to maintain net mass balance between the surface and atmosphere.

If atmospheric dynamics were dictated by both vapour pressure gradients and Coriolis forces, then it would  lead to the formation of Rossby waves and possibly weather systems. The ocean flow is slower than the atmosphere, so it would be even more affected by Coriolis forces. Ultimately, a Global Circulation Model (GCM) is required to accurately infer atmospheric dynamics that arise from the planet's rotation. However, this is difficult for a lava planet because the GCM would need to handle supersonic winds in an atmosphere that fully collapses far away from the sub-solar point.

Another approximation is the fixed surface temperature which allowed for the energy of the surface and atmosphere to be decoupled. This adds significant stability to the model especially as supersonic flow is guaranteed. Coupling the energy budgets of the surface and atmosphere in post-processing can help to estimate the actual surface temperature. We calculated how much energy it would take to heat up the advected ocean mass to the equilibrium surface temperature from the model’s results. This is done by taking the bulk transported mass per $\theta$ and multiplying it with the heat capacity and the temperature gradient. This energy peaks at ~10$^2$ W/m$^2$ which is minuscule compared to other energy contributions. 

Latent heat, may be a more significant source of energy for the surface. But the maximum latent heat produced by the evaporation of silica ($\sim$10$^2$ W/m$^2$), silicon monoxide, and sodium (both $\sim$10$^5$ W/m$^2$) are at least an order of magnitude smaller than the instellation ($\sim$10$^6$ W/m$^2$). This suggests that latent heat would barely affect the surface temperature, in agreement with \cite{castan2011atmospheres} and \cite{kite2016atmosphere}. It remains to be seen whether the energy of dissociation (e.g., SiO$_2$ -> SiO + O) is as important for lava planets as it has proven to be for ultra hot Jupiters (\citealp{bell2018increased}; \citealp{tan2019atmospheric}; \citealp{mansfield2020evidence}).

Another important limitation if our model is the lack of consideration for radiative processes in the atmosphere. Although the work of \cite{ito2015theoretical} suggests that there should be strong radiation absorption in the upper atmosphere, their results are sensitive to specific atmospheric compositions, stellar spectra in the ultraviolet, and atomic/molecular line lists, all of which are difficult to know for exoplanets.

The composition of the atmosphere is also very idealized as we ignore chemical processes where SiO$_2$ or Na$_2$O dissociate to form O and O$_2$, both of which are more volatile than SiO$_2$ but less volatile than Na. This has huge implications on transit spectroscopy as each molecule has its own distinct spectral features. Even in our idealized case where we focused on evaporation and condensation at great length, we neglected macro-scale meteorological processes, chiefly cloud formation. This also greatly affects observations as clouds can trap heat via longwave radiation absorption but reflect shortwave radiation.

\section{Conclusions}

K2-141b is the best target to study an exotic situation where a planet's interior, ocean, and atmosphere are compositionally similar. It will soon be observable with the James Webb Space Telescope and with ground-based telescopes equipped with high-resolution spectrographs. Although previous studies focused on more volatile species such as Na or K (\citealp{castan2011atmospheres,kite2016atmosphere}), we have shown that the significantly less volatile SiO$_2$ atmosphere may be easier to observe due to the extreme geometry of this system. Transit spectroscopy of K2-141b is possible due to its proximity to its star which make dayside regions accessible at the start and end of transit, and leads to  large acceleration throughout transit. Doppler shifts in transmission spectra could confirm the km/s wind speed of the atmosphere. Spectral phase measurements should be able to distinguish between the planetary surface and the somewhat cooler atmosphere. Since radiative transfer simulations predict a dayside temperature inversion \citep{ito2015theoretical}, phase curves may show a transition from inverted to non-inverted spectra.

Our calculated return flow of SiO$_2$ through the magma ocean is slow ($10^{-4}$ m/s), so mass balance can be easily achieved. This supports a steady-state silicate flow whereas the volatile Na atmosphere would be limited by the return flow through the magma ocean. Although K2-141b is an especially good target for atmospheric observation, our results regarding atmospheric dynamics, replenishment via the magma ocean, and transit geometry may apply to other lava planets.

\section*{Acknowledgements}

We would like to thank Raymond Pierrehumbert for his comments on the pre-submission draft as well as mentorship in hydrodynamical modelling. This work was made possible by the Natural Science and Engineering Research Council (NSERC) of Canada's Collaborative Research and Training Experience Program (CREATE) for Technology for Exo-planetary Sciences (Teps). This work is also made possible by Mathematics of Information Technology and Complex Systems (MITACS) Globalink Intership program. NBC acknowledges support from the McGill Space Institute (MSI) and
l'Institut de recherche sur les exoplan\`etes (iREx)

\section*{Data availability}

The data underlying this article will be shared on reasonable request to the corresponding author.


\bibliographystyle{mnras}
\bibliography{HotXplanet} 


\appendix
\section{SiO$_2$ phase diagram}
\label{App_SiO2phase}

\begin{figure}
	\centering	
	\includegraphics[width=\columnwidth,height=\textheight,keepaspectratio]{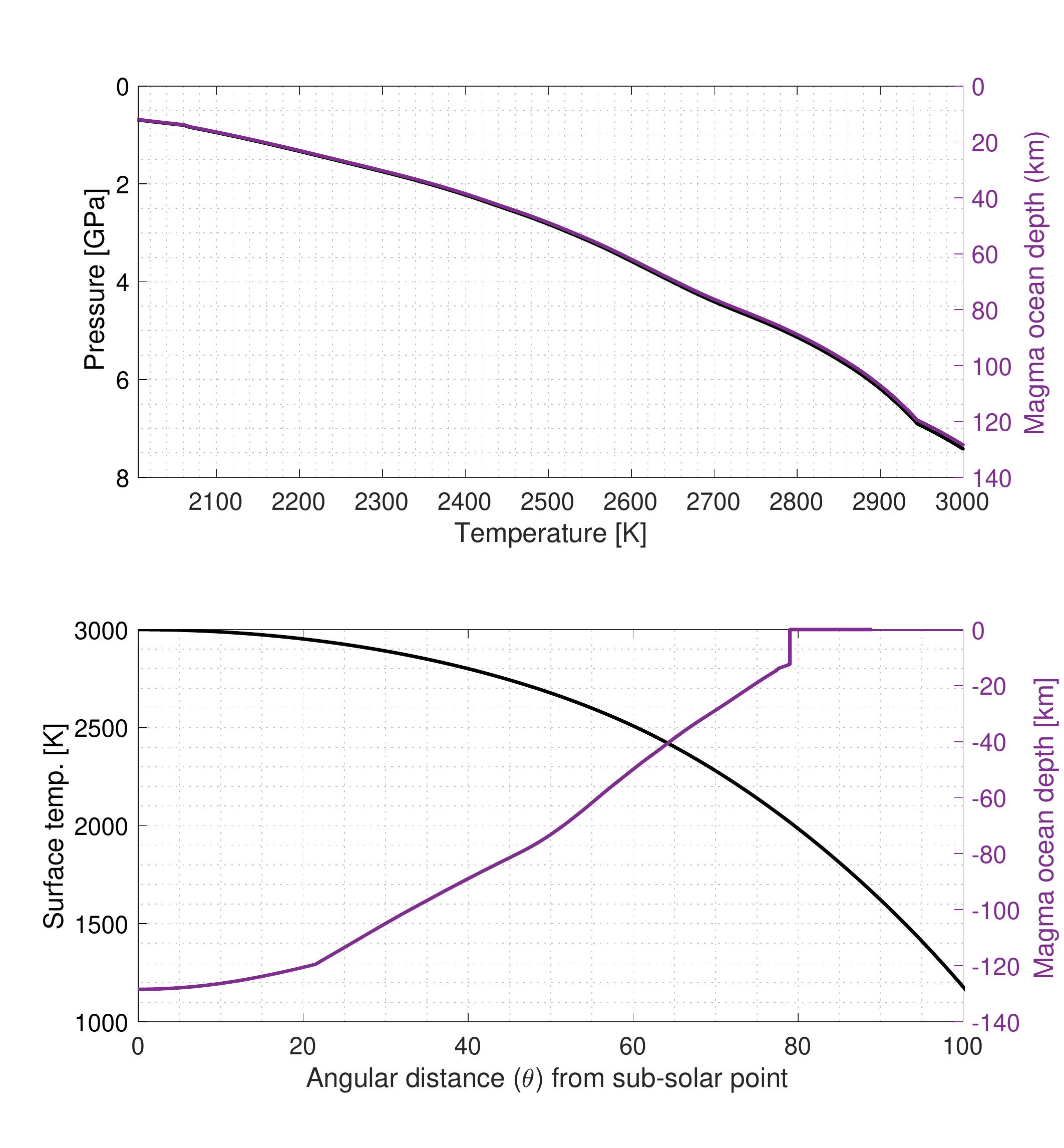}
	\caption{Top: phase diagram (black) and depth (purple) of a SiO$_2$ magma ocean. Left axis shows the pressure while the right shows the depth at which the phase will change between solid and liquid given the temperature and pressure. The inflection points reflect transitions between different SiO$_2$ crystal structures which are, in increasing pressure: cristobalite, $\beta$-quartz, coesite, and stishovite \citep{swamy1994thermodynamic}. Bottom: surface temperature (left/black) and magma ocean depth (right/purple) as a function of angular distance from subsolar point.}
	\label{fig_phasediagram}	
\end{figure}

See Fig. \ref{fig_phasediagram}.

\section{Gas specific parameters}
\label{App_Gas}

This subsection includes molecule specific parameters of Na, SiO, and SiO$_2$. The mass per molecule for the three constituents are: $m_{Na} = 3.82\times10^{-26}$ kg/molecule, $m_{SiO} = 7.32\times10^{-26}$ kg/molecule, and $m_{SiO_2} = 9.98\times10^{-26}$ kg/molecule. With this, you can compute the respective gas constant $R$ with $R = k/m$ where $k$ is the Boltzmann constant.

The heat capacity $C_p$ of Na, SiO, and SiO$_2$ is 904 J/K/kg \citep{castan2011atmospheres}, 851 J/K/kg (evaluated from the Shomate equation at 2400 K provided by \citealp{chase1998nist}), and 1428 J/K/kg \citep{lim2002vaporization} respectively. With this, $\beta$ can be calculated with the gas constant $R$ where $\beta = R/(R + C_p)$.

The final component is the saturated vapour pressure approximated from the model of \cite{schaefer2009chemistry} where: $P_v^{Na} [Pa] = 10^{9.6}e^{-38000/T_s}$, $P_v^{SiO} [Pa] = 6.16\times10^{13} e^{-69400/T_s}$, and $P_v^{SiO_2} [Pa] = 1.07\times10^{12} e ^{-66600/T_s}$.

\section{Calculating the fluxes}
\label{App_flux}

\subsection*{Mass}
$E$ is the mass flux from Eq. \ref{eq_mass}, which is essentially calculated by the evaporation/sublimation of the surface through the difference between the saturated vapour pressure and the atmospheric pressure:

\begin{equation}
E = \frac{(P_v-P)v_c}{\sqrt{2\pi} \ m \ R \ T_s}.
\label{eq_E}
\end{equation}

the kinetic speed of sublimation/evaporation is denoted by $v_c$, which for any species sublimating into a near vacuum, is $\sqrt{kT_s/m}$, where $k$ is the Boltzmann's constant.

\subsection*{Momentum}

The momentum flux $\tau$ is calculated by the following:

\begin{equation}
\tau = \rho_s (-V \ w_a),
\label{eq_tau}
\end{equation}

\noindent where $\omega_a \ [m/s]$, the transfer coefficient as defined by \cite{ingersoll1985supersonic}, is dependent on two velocity values: $V_e$ and $V_d$. There is a negative sign to denote that this force always acts against the flow. The vapour density, $\rho$ is determined via the ideal gas law. $V_e$ represents the mean flow and is calculated by: $V_e= m \ E/\rho_s$. $V_d$ represents the effects of eddies which is calculated by: $V_d = V_*^2/V$ where $V_*$ is the friction velocity; this must be calculated iteratively from:

\begin{equation}
V_{*,n+1} = \frac{V}{2.5 \ log(9 \ V_{*,n} \ H \ \rho_s/(2\eta) )}
\label{eq_V*}
\end{equation}

\noindent where $H$ is the scaled height, $\eta$ is the dynamic viscosity calculated by: $\eta(T_s) \ [kg \ m^{-1} \ s^{-1}] = 1.8\times10^{-5}(T_s/291)^{3/2}(411/(T_s + 120))$ \citep{castan2011atmospheres}.

\subsection*{Energy}

The energy flux $Q$, a combination of enthalpy at surface and top of boundary layer as well as kinetic energy, is calculated by the following:

\begin{equation}
Q = \rho_s(w_s \ C_p \ T_s - w_a \ (V^2/2 + C_p \ T))
\label{eq_Q}
\end{equation}

\noindent where $w_s$ is the other transfer coefficient defined by \cite{ingersoll1985supersonic}. Both $w_a$ and $w_s$ are evaluated from Equation (\ref{eq_wa1}) to (\ref{eq_ws2}).

\begin{equation}
w_a = \frac{V_e^2 - 2V_d \ V_e + 2V_d^2}{-V_e + 2 V_d} \ , V_e < 0
\label{eq_wa1}
\end{equation}

\begin{equation}
w_a = \frac{2V_d^2}{V_e + 2 V_d} \ , V_e \geq 0
\label{eq_wa2}
\end{equation}

\begin{equation}
w_s = \frac{2V_d^2}{-V_e + 2 V_d} \ , V_e \leq 0
\label{eq_ws1}
\end{equation}

\begin{equation}
w_s = \frac{V_e^2 - 2V_d \ V_e + 2V_d^2}{V_e + 2 V_d} \ , V_e > 0
\label{eq_ws2}
\end{equation}

\section{Numerically solving the system of equations}
\label{App_solve}

The state variables $(P,V,T)$ can be calculated given initial values at the boundary (subsolar and antisolar point) by integration of fluxes in the right-hand side of Eq. (\ref{eq_mass}) - (\ref{eq_en}). If we treat these equation as a single equation, we get:

\begin{equation}
\frac{d}{d\theta}(f_{state}(\theta)) = f_{fluxes}(\theta)
\end{equation}

The state variable can be calculated at the next spatial step (the change in angular distance $\Delta\theta$) through finite differences:

\begin{equation}
f_{state}(\theta + \Delta\theta) = f_{fluxes}(\theta) \times \Delta\theta + f_{state}(\theta)
\end{equation}

We solve for a solution through finite-differences using a Runge-Kutta scheme to the second order accuracy. After acquiring $f_{state}$, the state variables can be solved arithmetically but due to the quadratic nature of $V$ in the equations, two branches can be extracted corresponding to the flow being either subsonic or supersonic. Although, a higher order scheme could remedy the sensitivity of modelling fluid transitioning from subsonic to supersonic flow, the calculations to extract the state variables from Eq. (\ref{eq_mass}) - (\ref{eq_en}) would take longer. Therefore, we elect to employ other schemes to control the instability inherent in the model, as described below.

We use the shooting method where we guess the initial $P(\theta=0)$ and find a solution that fits the boundary conditions: $V$ is 0 at $\theta=0\degree$ and $\theta=180\degree$, $T(\theta=0) = T_{ss}$. The instability of the non-linear system of equations can lead to solutions either exponentially expanding or decaying. The goal, therefore, is to solve for a solution that exists in between the bounds of solutions that blew up (upper) or decayed to 0 (lower) exponentially. Conditions that define whenever a solution expand or decay exponentially is explicitly described by \cite{ingersoll1985supersonic}.

Solving for velocity will yield two solutions, one for subsonic flow and the other for supersonic flow.
Getting to the critical point where the flow transition between different subsonic and supersonic regimes can be difficult. Therefore, we dynamically increase the spatial resolution whenever the upper bound and lower bound solution start to diverge past a set parameter. A solution close to the "true" solution (which lies in between the upper and lower bound) will approach Mach 1 more reliably, after which we extrapolate the flow into the supersonic regime. This extrapolation process is done by linearizing the state variables with respect to $\theta$ until Mach 1 is achieved. The flow is then calculated using the supersonic branch of the solutions.


\section{Observational inference calculations}
\label{App_obs}

The predicted blackbody radiation of K2-141b required to produce Fig. \ref{fig_emiss} is calculated using the methods from \cite{cowan2008inverting} and \cite{cowan2011statistics}. The first step is to convert temperature, $T$ to blackbody radiative intensity, $I$ [W m$^{-2}$ m$^{-1}$ sr$^{-1} $] at wavelength $\lambda$:

\begin{equation}
I = \frac{2 h c^2}{\lambda^5} \frac{1}{e^{h c/(\lambda k T)} - 1},
\end{equation}

\noindent where $h$ is the Planck constant, $c$ is the speed of light, and $k$ is the Boltzmann constant. The next step is to create an intensity map with latitude ($\alpha$) and longitude ($\phi$) from angular distance ($\theta$) from the subsolar point using the Spherical Pythagorean Theorem:

\begin{equation}
I(\alpha,\phi) = I(\theta = \cos^{-1}[\cos(\alpha)\cos(\phi)]).
\end{equation}

 Now that each latitude ($\alpha$) and longitude ($\phi$) coordinate has a radiative intensity, we then integrate the intensity across the hemisphere visible to the observer. Assuming an edge-on viewing geometry where the observer is on the planet's orbital plane, the observed radiative flux, $F$, is calculated by:

\begin{equation}
F(\xi) = \int_{0}^{\pi} \int_{\xi-\pi/2}^{\xi+\pi/2} I(\alpha,\phi) \sin^2 \alpha \cos(\phi-\xi) \ d\alpha \ d\phi,
\end{equation}

\noindent where $\xi$ is the planet's position around the orbit. The stellar flux at the corresponding wavelength is calculated in the exact same way but with a uniform temperature of 4599 K \citep{malavolta2018ultra} and scaled to the ratio of $(r_{star}/r_{planet})^2$. This therefore yields the ratio of planetary to stellar flux at that point as a function of the planet's orbital motion about its star, the so-called phase curve. 

Brightness temperature $T_b$ is calculated from \citep{cowan2011statistics}:

\begin{equation}
T_b = \frac{h c}{\lambda k}\bigg[ \log \bigg(1+\frac{e^{hc/\lambda k T_b^*} -1}{\psi} \bigg) \bigg]^{-1},
\end{equation}

\noindent where $T_b^*$ is the star's effective temperature (4599 K) and $\psi$ is the radiative flux ratio between the planet and the star.

\bsp	
\label{lastpage}

\end{document}